\documentstyle[12pt]{article}

\newfont{\goth}{cmbxti10 scaled\magstep1}
\newfont{\gothi}{cmbxti10}

\newcommand{\smin}{\,\raisebox{0.06em}{${\scriptstyle \in}$}\,}
\newcommand{\smsubset}{\,\raisebox{0.06em}{${\scriptstyle \subset}$}\,}

\newcommand{\gotg}{\mbox{\goth g}}

\begin{document}
\title{Current Algebra of Classical Non-Linear Sigma Models}
\author{M. Forger \\
        J. Laartz \thanks{Address after September 1, 1991:
                   Department of Mathematics,
                   Harvard University,
                   1 Oxford Street,
                   Cambridge MA 02138 / USA}\\
        U. Sch\"aper}

 \date{Fakult\"at f\"ur Physik der Universit\"at Freiburg, \\
      Hermann-Herder-Str. 3, D-7800 Freiburg / FRG  \vspace{2em}}

\maketitle
\thispagestyle{empty}
\begin{abstract}
\noindent
The current algebra of classical non-linear sigma models on arbitrary
Riemannian
manifolds is analyzed. It is found that introducing, in addition to the Noether
current $j_\mu$ associated with the global symmetry of the theory, a composite
scalar field $j$, the algebra closes under Poisson brackets.
\end{abstract}

\vfill

 \begin{flushright}
 \parbox{12em}
 { \begin{center}
 University of Freiburg \\
 THEP 91/10 \\
 August 1991
 \end{center} }
 \end{flushright}

\newpage
\setcounter{page}{1}

It is well known that for the classical non-linear sigma models, the Poisson
brackets between the Noether currents associated with the global symmetry of
the theory involve Schwinger terms which, in general, are field dependent.
Explicit expressions have been written down, e.g., for the principal chiral
models (sigma models on compact Lie groups) \cite{FaTa,Mai1} and for the
spherical models ($O(N)$-invariant sigma models on spheres) \cite{Mai2,DEM1},
but the structure of the resulting algebra has, to our knowledge, not yet
been analyzed in full generality.

In the following, we want to show that, for classical non-linear sigma models
on arbitrary Riemannian manifolds, introducing the coefficient of the Schwinger
term appearing in the Poisson bracket between the time component and the space
component of the Noether current $j_\mu$ as a new composite field $j$ leads to
an algebra which closes under Poisson brackets: this is what we propose to call
the current algebra for these models.

To this end, consider the classical two-dimensional non-linear sigma model on
an arbitrary Riemannian manifold $M$ with metric $g$. The configuration space
of the theory consists of (smooth) maps $\varphi$ from a fixed two-dimensional
Lorentz manifold $\Sigma$ (typically two-dimensional Minkowski space) to $M$,
while the corresponding phase space consists of pairs $(\varphi,\pi)$ of
fields,
$\pi$ being a smooth section of the pull-back $\, \varphi^\ast (T^\ast M) \,$
of the cotangent bundle of $M$ to $\Sigma$ via $\varphi$. In terms of local
coordinates $u^i$ on $M$, $\varphi$ and $\pi$ are represented by multiplets
of ordinary functions $\varphi^i$ and $\pi_i$ on $\Sigma$; then the action
reads
\begin{equation}
S~=~{\textstyle{1\over 2}} \int d^2\! x \; g_{ij}(\varphi) \,
    \partial^\mu \varphi^i \, \partial_\mu\varphi^j~~~,            \label{eq:S}
\end{equation}
and the canonical Poisson brackets are
\begin{eqnarray}
 \{ \varphi^i(x)\, , \varphi^j(y) \} = 0 &,& \{ \pi_i(x) , \pi_j(y) \} = 0~~~,
                                                              \nonumber \\[1mm]
 \{ \varphi^i(x)\, , \pi_j(y) \} &=& \delta^i_j \, \delta(x-y)~~~.
                                                                 \label{eq:CCR}
\end{eqnarray}
Denoting the time derivative by a dot and the spatial derivative by a prime, we
have
\begin{equation}
 \pi_i~=~g_{ij}(\varphi) \, \dot{\varphi}^j~~~.
\end{equation}
Under local coordinate transformations $~u^i \rightarrow u^{\prime k} \,$, the
component fields $\varphi^i$, $\partial_\mu \varphi^i$ and $\pi_i$ transform
according to
\begin{equation}
 \varphi^i~\rightarrow~\varphi'^k~~~,~~~
 \partial_\mu \varphi^i~\rightarrow~
 \partial_\mu \varphi'^k = {\partial\varphi'^k \over \partial\varphi^i} \,
                            \partial_\mu \varphi^i~~~,~~~
 \pi_i~\rightarrow~\pi'_k = {\partial\varphi^i \over \partial\varphi'^k} \,
                            \pi_i~~~,
\end{equation}
from which it can be checked that the action (\ref{eq:S}) and the canonical
commutation relations (\ref{eq:CCR}) are invariant.

Next, let $G$ be a (connected) Lie group acting on $M$ by isometries. Then
every generator $X$ in the Lie algebra $\gotg\,$ of $G$ is represented by a
fundamental vector field $X_M$ on $M$, given by
\begin{equation}
 X_M(m)~=~{d\over dt} \, (\exp(tX) \cdot m) \, \Big|_{t=0}~~~,  \label{eq:FVF1}
\end{equation}
obeying the representation condition
\[
 [X,Y]_M \, = \mbox{} - [X_M,Y_M]~~~;
\]
moreover, all these vector fields are Killing vector fields. As usual,
$G$-invariance of the action (\ref{eq:S}) leads to a conserved Noether
current $j_\mu$ with values in $\gotg^\ast$ (the dual of $\gotg\,$).
Explicitly, for $\, X\smin\gotg\,$,
\begin{equation}
 \langle j_\mu , X \rangle~=~- \, g_{ij}(\varphi) \, \partial_\mu \varphi^i \,
                                  X_M^j(\varphi)~~~.             \label{eq:NC1}
\end{equation}
On the other hand, we define a scalar field $j$ with values in the
symmetric tensor product of ${\goth g}^\ast$ with itself by setting,
for $\, X,Y\smin\gotg\,$,
\begin{equation}
 \langle j , X\otimes Y \rangle~=~g_{ij}(\varphi) \, X_M^i(\varphi) \,
                                 Y_M^j(\varphi)~~~.              \label{eq:AF1}
\end{equation}
In terms of an arbitrary basis ($T_a$) of $\gotg\,$, with structure
constants $f_{ab}^c$ defined by \mbox{$[T_a,T_b] = f_{ab}^c T_c \;$},
and the corresponding dual basis $(T^a)$ of $\gotg^\ast$, we write
\begin{equation}
 j_\mu~=~j_{\mu,a} \, T^a~~~,~~~j~=~j_{ab} \; T^a\otimes T^b~~~.
\end{equation}
Then the current algebra takes the following form:
\begin{eqnarray}
 \{ j_{0,a}(x) , j_{0,b}(y) \}
 &=& - \, f_{ab}^c \, j_{0,c}(x) \, \delta(x-y)~~~,      \label{eq:CA1} \\[1mm]
 \{ j_{0,a}(x) , j_{1,b}(y) \}
 &=& - \, f_{ab}^c \, j_{1,c}(x) \, \delta(x-y) \,
     + \, j_{ab}(y) \, \delta^\prime(x-y)~~~,            \label{eq:CA2} \\[1mm]
 \{ j_{1,a}(x) , j_{1,b}(y) \} &=& \, 0~~~,              \label{eq:CA3} \\[2mm]
 \{ j_{0,a}(x) , j_{bc}(y) \}
 &=& - \left( f_{ab}^d \, j_{cd}(x) + f_{ac}^d \, j_{bd}(x) \right)
                                    \delta(x-y)~~~,      \label{eq:CA4} \\[1mm]
 \{ j_{1,a}(x) , j_{bc}(y) \} &=& \, 0~~~,               \label{eq:CA5} \\[2mm]
 \{ j_{ab}(x) , j_{cd}(y) \} &=& \, 0~~~.                \label{eq:CA6}
\end{eqnarray}

For the proof, we note first of all that the Poisson brackets
(\ref{eq:CA3}), (\ref{eq:CA5}) and (\ref{eq:CA6}) vanish because
$j_{1,a}$ and $j_{ab}$ depend only on the $\varphi^i$ but not on
the $\pi_i$. The Poisson bracket of two $j_0$'s is:
\begin{eqnarray*}
\lefteqn{\{ \langle j_0(x) , X \rangle \, , \, \langle j_0(y) , Y \rangle \}~~
 =~~\{ \pi_i(x)\, X_M^i(\varphi(x)) \, , \, \pi_j(y)\, Y_M^j(\varphi(y)) \}}
                                                         \hspace{1.5cm} \\[1mm]
 &=& \pi_i(x) \left( \partial_j X_M^i(\varphi(x)) \, Y_M^j(\varphi(x))
                   - X_M^j(\varphi(x)) \, \partial_j Y_M^i(\varphi(x)) \right)
     \delta(x-y)                                                      \\[0.5mm]
 &=& \pi_i(x) \, [X,Y]_M^i(\varphi(x)) \, \delta(x-y)                   \\[1mm]
 &=& \mbox{} - \langle j_0(x) , [X,Y] \rangle \, \delta(x-y)~~~.
\end{eqnarray*}
The Poisson bracket of a $j_0$ with a $j_1$, however, is more complicated:
\begin{eqnarray*}
\{ \langle j_0(x) , X \rangle \, , \, \langle j_1(y) , Y \rangle \}
 &=& \{ \pi_i(x) \, X_M^i(\varphi(x)) \, , \,
        g_{jk}(\varphi(y)) \, \varphi^{\prime j}(y) \, Y_M^k(\varphi(y)) \}
                                                                        \\[2mm]
 &=& X_M^i(\varphi(x)) \, g_{ik}(\varphi(y)) \, Y_M^k(\varphi(y)) \,
                       \delta^\prime(x-y)                             \\[0.5mm]
 & & \mbox{} - X_M^i(\varphi(x)) \, \partial_i g_{jk}(\varphi(x)) \,
               Y_M^k(\varphi(x)) \, \varphi^{\prime j}(x) \, \delta(x-y)
                                                                      \\[0.5mm]
 & & \mbox{} - X_M^i(\varphi(x)) \, g_{jk}(\varphi(x)) \,
               \partial_i Y_M^k(\varphi(x)) \, \varphi^{\prime j}(x) \,
               \delta(x-y)~~~.                                     \vspace{1mm}
\end{eqnarray*}
Using the identity
\[
 f(x) \, \delta^\prime(x-y)~=~f(y) \, \delta^\prime(x-y)
                            - f^\prime(x) \, \delta(x-y)~~~,
\]
together with the fact that $X_M$ is a Killing field, so
\[
 \partial_j X_M^i \; g_{ik}~=~- \, \partial_k X_M^i \; g_{ij}
                              - \partial_l g_{jk} \, X^l_M~~~,
\]
we get
\begin{eqnarray*}
\{ \langle j_0(x) , X \rangle \, , \, \langle j_1(y) , Y \rangle \}
 &=& X_M^i(\varphi(y)) \, g_{ik}(\varphi(y)) \, Y_M^k(\varphi(y)) \,
                       \delta^\prime(x-y)                             \\[0.5mm]
 & & \mbox{} + \partial_k X_M^i(\varphi(x)) \, g_{ij}(\varphi(x)) \,
          Y_M^k(\varphi(x)) \, \varphi^{\prime j}(x) \, \delta(x-y)   \\[0.5mm]
 & & \mbox{} - X_M^k(\varphi(x)) \, g_{ij}(\varphi(x)) \,
               \partial_k Y_M^i(\varphi(x)) \, \varphi^{\prime j}(x) \,
               \delta(x-y)                                              \\[1mm]
 &=& X_M^i(\varphi(y)) \, g_{ik}(\varphi(y)) \, Y_M^k(\varphi(y)) \,
                       \delta^\prime(x-y)                             \\[0.5mm]
 & & \mbox{} - g_{ij}(\varphi(x)) \, [X_M,Y_M]^i(\varphi(x)) \,
               \varphi^{\prime j}(x) \, \delta(x-y)                     \\[1mm]
 &=& \mbox{} - \langle j_1(x) , [X,Y] \rangle \, \delta(x-y) \;
             + \; \langle j(y) , X\otimes Y \rangle \, \delta^\prime(x-y)~~~.
\end{eqnarray*}
Finally, the Poisson bracket of a $j_0$ with a $j$ reads:
\begin{eqnarray*}
\{ \langle j_0(x) , X \rangle \, , \, \langle j(y) , Y\otimes Z \rangle \}
 &=& - \, \{ \pi_i(x) \, X_M^i(\varphi(x)) \, , \,
                         g(Y_M,Z_M)(\varphi(y)) \}                      \\[1mm]
 &=& {\bf L}_{X_M} (g(Y_M,Z_M))(\varphi(x)) \; \delta(x-y)              \\[1mm]
 &=& - \, \langle \, j(x) \, , \, [X,Y]\otimes Z + Y\otimes [X,Z] \,
                                                   \rangle \; \delta(x-y)
\end{eqnarray*}
(since $~{\bf L}_{X_M} g = 0 \,$, ${\bf L}$ being the Lie derivative).

The definition of the model and the derivation of the current algebra
(\ref{eq:CA1}-\ref{eq:CA6}) have here been given in the (gauge invariant)
local coordinate formulation, which is the only one available for general
Riemannian manifolds. For the special case where the target manifold is
Riemannian homogeneous space $\; M = G/H \,$, however, there is an
alternative (gauge dependent) formulation [5-8] % \cite{EF1,EF2,EF3,For}
in terms of fields $g$ with values in $G$, determined modulo fields $h$
with values in $H$ by the condition that $~\varphi = gH~$ (at least locally),
and it is instructive to rewrite the definition of the composite fields
$j_\mu$ and $j$ in this language. The assumption that $M$ is a Riemannian
homogeneous space implies that it can be written as the quotient space
of some (connected) Lie group $G$, with Lie algebra $\mbox{\goth g}\,$,
modulo some compact subgroup $\; H \smsubset G \,$, with Lie algebra
$\; \mbox{\goth h} \smsubset \mbox{\goth g} \,$, and that it is reductive.
This means that there exists an $\mbox{Ad}(H)$-invariant subspace
$\mbox{\goth m}\,$ of $\mbox{\goth g}\,$ such that $\mbox{\goth g}\,$ is
the (vector space) direct sum of $\mbox{\goth h}\,$ and $\mbox{\goth m}\,$:
\begin{equation}
 \mbox{\goth g}\, = \mbox{\goth h}\, \oplus \mbox{\goth m}~~~.    \label{eq:DD}
\end{equation}
The corresponding projections from $\mbox{\goth g}\,$ onto $\mbox{\goth h}\,$
along $\mbox{\goth m}\,$ and from $\mbox{\goth g}\,$ onto $\mbox{\goth m}\,$
along $\mbox{\goth h}\,$ will be denoted by $\, \pi_{\mbox{\gothi h}} \,$ and
$\, \pi_{\mbox{\gothi m}} \,$, respectively. Due to $\mbox{Ad}(H)$-invariance
of the direct decomposition (\ref{eq:DD}), we have the commutation relations
\begin{equation}
 [\mbox{\goth h}\, , \mbox{\goth h}\,] \smsubset \mbox{\goth h}~~~,~~~
 [\mbox{\goth h}\, , \mbox{\goth m}\,] \smsubset \mbox{\goth m}~~~.
                                                                 \label{eq:CR1}
\end{equation}
Moreover, we shall suppose that the $\mbox{Ad}(H)$-invariant positive
definite inner product $(.\, ,.)$ on $\mbox{\goth m}\,$, corresponding
to the given $G$-invariant Riemannian metric on $M$, is induced from an
$\mbox{Ad}(G)$-invariant non-degenerate inner product $(.\, ,.)$ on
$\mbox{\goth g}\,$, corresponding to a $G$-biinvariant pseudo-Riemannian
metric on $G$, so that the direct decomposition (\ref{eq:DD}) is orthogonal.
(This amounts essentially to requiring that $M$ be naturally reductive; we
refer to \cite{For} for a detailed discussion.) Now using this scalar product
on $\mbox{\goth g}\,$ to pull up and down Lie algebra indices, we can interpret
$j_\mu$ as a vector field with values in $\mbox{\goth g}\,$ and $j$ as a scalar
field with values in the space $~\mbox{End}(\mbox{\goth g}\,) \cong
\mbox{\goth g}\, \otimes \mbox{\goth g}^\ast~$ of endomorphisms of
$\mbox{\goth g}\,$ (i.e., of linear transformations on $\mbox{\goth g}\,$).
Then using that (\ref{eq:FVF1}) can be rewritten as
\begin{equation}
 X_M(gH)~=~g \, \pi_{\mbox{\gothi m}} (\mbox{Ad}(g)^{-1}X)~~~,  \label{eq:FVF2}
\end{equation}
and that the covariant derivative $\, D_\mu g \,$ of $g$ is defined by
\begin{equation}
 D_\mu g~=~g \, \pi_{\mbox{\gothi m}} (g^{-1} \, \partial_\mu g)~~~,
\end{equation}
we get
\begin{eqnarray*}
 (j_\mu , X) &=& \mbox{} - (\partial_\mu \varphi \, , \, X_M(\varphi))
 ~~=~~\mbox{} - (D_\mu g \, , \, g \, \pi_{\mbox{\gothi m}}
                                      (\mbox{Ad}(g)^{-1}X))           \\[0.1cm]
  &=& \mbox{} - (g^{-1}D_\mu g \, , \, \pi_{\mbox{\gothi m}}
                                       (\mbox{Ad}(g)^{-1}X))
 ~~=~~\mbox{} - (g^{-1}D_\mu g \, , \, \mbox{Ad}(g)^{-1}X)            \\[0.1cm]
  &=& \mbox{} - (D_\mu g \, g^{-1} , X)~~~,
\end{eqnarray*}
and
\begin{eqnarray*}
 (j(X),Y) &=& (X_M(\varphi) , Y_M(\varphi))
 ~~=~~(\pi_{\mbox{\gothi m}} (\mbox{Ad}(g)^{-1}X) \, , \,
       \pi_{\mbox{\gothi m}} (\mbox{Ad}(g)^{-1}Y))                    \\[0.1cm]
  &=& ((\pi_{\mbox{\gothi m}} \, \mbox{Ad}(g)^{-1}) X \, , \,
                                 \mbox{Ad}(g)^{-1}Y)
 ~~=~~((\mbox{Ad}(g) \, \pi_{\mbox{\gothi m}} \, \mbox{Ad}(g)^{-1}) X \, , \,
                                                  Y)~~~,
\end{eqnarray*}
i.e.,
\begin{equation}
 j_\mu~=~\mbox{} - D_\mu g \, g^{-1}~~~,                         \label{eq:NC2}
\end{equation}
and
\begin{equation}
 j~=~\mbox{Ad}(g) \, \pi_{\mbox{\gothi m}} \, \mbox{Ad}(g)^{-1}~~~.
                                                                 \label{eq:AF2}
\end{equation}
The formula for the Noether current $j_\mu$ is well known [5-8],
% \cite{EF1,EF2,EF3,For},
while the formula for the new field $j$ simply states that it is conjugate
to the projector $\, \pi_{\mbox{\gothi m}} \,$. It should also be noted that
the two fields are not independent. Thus for example, we have the following
identity expressing the derivatives of $j$ in terms of $j$ and $j_\mu$:
\begin{equation}
 \partial_\mu j~=~[ \, j \, , \, \mbox{ad}(j_\mu) \, ]
               ~=~j \, \mbox{ad}(j_\mu) \, - \, \mbox{ad}(j_\mu) \, j~~~.
                                                                 \label{eq:AF3}
\end{equation}
If $\; M = G/H \;$ is not only a homogeneous space but even a symmetric one,
so that in addition to (\ref{eq:CR1}), we have the commutation relation
\begin{equation}
 [\mbox{\goth m}\, , \mbox{\goth m}\,] \smsubset \mbox{\goth h}~~~,
                                                                 \label{eq:CR2}
\end{equation}
then we get an additional algebraic identity between $j$ and $j_\mu$:
\begin{equation}
 \mbox{ad}(j_\mu)~=~j \, \mbox{ad}(j_\mu) \, + \, \mbox{ad}(j_\mu) \, j~~~.
                                                                 \label{eq:AF4}
\end{equation}
This case is of particular importance since it is precisely the non-linear
sigma models on Riemannian symmetric spaces which, in two dimensions, are
integrable field theories. The implications of the current algebra derived
above for the canonical structure of these integrable models will be
discussed in a separate paper \cite{BFLS}.

Incidentally, the derivation of the current algebra (\ref{eq:CA1}-\ref{eq:CA6})
given above is not restricted to the two-dimensional case but is valid for
Lorentz manifolds $\Sigma$ of arbitrary dimension. Indeed, passing to higher
dimensions means that the spatial variables $x,y,\ldots$ become vectors
$\vec{x},\vec{y},\ldots\,$, and the only changes required are that the spatial
derivative $.'$ is replaced by the gradient $\vec{\nabla}$ and $j_1$ by
$\vec{\jmath}$.

To conclude, we consider two examples which have been discussed in the
literature before, namely the principal chiral models \cite{FaTa,Mai1}
and the so-called $O(N)$-models \cite{Mai2,DEM1}.

\noindent
In the principal chiral models, the target space is a compact Lie group $G$,
equipped with a $G$-biinvariant metric $(.\, ,.)$. The symmetry group of the
theory is now the direct product $\; G_L \times G_R \;$ of two copies $G_L$
and $G_R$ of $G$, acting on $G$ by left and right translations, respectively.
Then $G$ becomes a Riemannian symmetric space by identifying it with the
quotient $\; G \times G / \Delta G \,$, $\Delta G$ being the diagonal subgroup.
Correspondingly, the fields $\varphi$ and $g$ used before are now denoted by
$g$ and $(g_1,g_2)$, respectively, and the formula $~\varphi = gH~$ becomes
$~g = g_1 g_2^{-1}$. With this notation, the Noether current $j_\mu$, having
values in $~\gotg_L \oplus \gotg_R \,$, splits into two components,
$~j_\mu = (j_\mu^L,j_\mu^R) \,$, with
\begin{equation}
 j_\mu^L~=~- \, {\textstyle{1\over 2}} \, \partial_\mu g \, g^{-1}~~~,~~~
 j_\mu^R~=~+ \, {\textstyle{1\over 2}} \, g^{-1} \, \partial_\mu g~~~,
\end{equation}
cf.~\cite{EF1}. Similarly, the scalar field $j$, having values in
$\; \mbox{End}(\gotg_L \oplus \gotg_R) \,$, can be written as a
$(2 \times 2)$-block matrix:
\begin{equation}
 j~=~{\textstyle{1\over 2}} \;
     \left( \begin{array}{cc}
             1 & - \mbox{Ad}(g) \\
             - \mbox{Ad}(g)^{-1} & 1
            \end{array} \right)~~~.
\end{equation}

\noindent
In the so called $O(N)$-models, we have $~M = S^{N-1} \, ,~G = SO(N) \,$.
The Noether current $j_\mu$, having values in $so(N)$, is given by the
well-known formula
\begin{equation}
 j_\mu~=~\varphi \, \partial_\mu\varphi^T \,
               - \, \partial_\mu \varphi \, \varphi^T~~~.
\end{equation}
Similarly, the scalar field $j$, having values in $\; \mbox{End}(so(N)) \,$,
is found to act on $\, X\smin so(N) \,$ according to
\begin{equation}
 j(X)~=~\varphi \varphi^T X \, + \, X \, \varphi \varphi^T~~~.
\end{equation}
(The proof is most conveniently performed by using that
$~\pi_{\mbox{\gothi m}} = {1\over 2} (1 - \sigma) \,$, where
\mbox{$\sigma = \mbox{Ad}(\theta)~$} with
$~\theta = \mbox{diag}(1,-1,\ldots\,,-1) \,$,
so $~g\theta g^{-1} = 2\varphi\varphi^T - 1 \,$.)

{\bf Acknowledgement:} We would like to thank M.~Bordemann for fruitful
discussions and C.~Emmrich for a careful reading of the manuscript.


\begin{thebibliography}{9}

\bibitem{FaTa} L.D.~Faddeev and L.A.~Takhtajan: {\em Hamiltonian Methods in the
 Theory of Solitons}, Springer-Verlag, Berlin 1987.

\bibitem{Mai1} J.-M.~Maillet: {\em Hamiltonian Structures for Integrable
 Classical Theories from Graded Kac-Moody Algebras},
 Phys.~Lett.~{\bf 167 B} (1986) 401-405.

\bibitem{Mai2} J.-M.~Maillet: {\em Kac-Moody Algebra and Extended Yang-Baxter
 Relations in the $O(N)$ Non-Linear Sigma Model},
 Phys.~Lett.~{\bf 162 B} (1985) 137-142.

\bibitem{DEM1} H.J.~de Vega, H.~Eichenherr and J.-M.~Maillet: {\em Classical
 and Quantum Algebras of Non-Local Charges in Sigma Models},
 Commun.~Math.~Phys.~{\bf 92} (1984) 507-524.

\bibitem{EF1} H.~Eichenherr and M.~Forger: {\em On the Dual Symmetry of the
 Nonlinear Sigma Models}, Nucl.~Phys.~{\bf B 155} (1979) 381-393.

\bibitem{EF2} H.~Eichenherr and M.~Forger: {\em More about Nonlinear Sigma
 Models on Symmetric Spaces}, Nucl.~Phys.~{\bf B 164} (1980) 528-535 \&
 {\bf B 282} (1987) 745-746 (erratum).

\bibitem{EF3} H.~Eichenherr and M.~Forger: {\em Higher Local Conservation Laws
 for Nonlinear Sigma Models on Symmetric Spaces}, Commun.~Math.~Phys.~{\bf 82}
 (1981) 227-255.

\bibitem{For} M.~Forger: {\em Nonlinear Sigma Models on Symmetric Spaces}. In:
 {\em Nonlinear Partial Differential Operators and Quantization Procedures},
 proceedings, Clausthal, Germany 1981, eds: S.I.~Andersson and H.D.~Doebner;
 Lecture Notes in Mathematics {\bf 1037}, Springer Verlag 1983.

\bibitem{BFLS} M.~Bordemann, M.~Forger, J.~Laartz and U.~Sch\"aper:
 {\em The Lie-Poisson Structure of Integrable Classical Non-Linear Sigma
 Models}, University of Freiburg preprint THEP 91/11.

\end{thebibliography}
\end{document}